\documentclass[aps, pra, twocolumn, amsmath, amssymb, superscriptaddress, nofootinbib]{revtex4-1}

\makeatletter
\def\p@subsection{}
\def\p@subsubsection{}
\makeatother

\usepackage[utf8]{inputenc}
\usepackage{graphicx}
\usepackage{units}
\usepackage{color}
\usepackage[pdftex, colorlinks=true, linkcolor=myblue, citecolor=myblue,
urlcolor=myblue]{hyperref}
\usepackage{times}
\usepackage{comment}
\usepackage{graphicx}
\usepackage{amsmath, calc}
\usepackage{mathrsfs}
\usepackage{amsfonts}
\usepackage{amssymb}
\usepackage{bm}
\usepackage{bbm}
\usepackage{color}
\usepackage{array}
\usepackage{units}
\usepackage{tabstackengine}
\stackMath

\usepackage{natbib}
\setcitestyle{citesep={;}}
\setcitestyle{round}

\definecolor{myblue}{rgb}{0,0,1}
\definecolor{myred}{rgb}{1,0,0}

\begin{document}


\title{Quantum confinement in Dirac-like nanostructures}


\author{C. A. Downing}
\email{c.a.downing@exeter.ac.uk} 
\affiliation{Department of Physics and Astronomy, University of Exeter, Exeter EX4 4QL, United Kingdom}

\author{M. E. Portnoi}
\email{m.e.portnoi@exeter.ac.uk} 
\affiliation{Department of Physics and Astronomy, University of Exeter, Exeter EX4 4QL, United Kingdom}



\begin{abstract}
\noindent \textbf{Abstract}\\
In Westminster Abbey, in a nave near to Newton's monument, lies a memorial stone to Paul Dirac. The inscription on the stone includes the relativistic wave equation for an electron: the Dirac equation. At the turn of the 21st century, it was discovered that this eponymous equation was not simply the preserve of particle physics. The isolation of graphene by Andre Geim and Konstantin Novoselov in Manchester led to the exploration of a novel class of materials – Dirac materials - whose electrons behave like Dirac particles. While the mobility of these quasi-relativistic electrons is attractive from the perspective of potential ultrafast devices, it also presents a distinct challenge: how to confine Dirac particles so as to avoid making inherently leaky devices? Here we discuss the unconventional quantum tunnelling of Dirac particles, we explain a strategy to create bound states electrostatically, and we briefly review some pioneering experiments seeking to trap Dirac electrons.
\\
\\
\noindent \textbf{Keywords}\\
Berry phase, chirality, Dirac equation, graphene, Klein tunnelling, linear spectrum, low-dimensional materials, Maxwell's fish-eye lens, nanomaterials, quantum confinement, quantum scattering, quantum transport, quasi-relativistic phenomena, zero bandgap semiconductors
\end{abstract}


\maketitle



\noindent \textbf{Key points}
\begin{itemize}
    \item the Dirac equation has utility in condensed matter physics, where it can describe the electrons in materials such as graphene
  \item  confinement in Dirac materials is notoriously difficult due to the Klein paradox (which arises during quantum tunnelling of quasi-relativistic particles)
  \item unexpectedly, bound states may form at the Dirac point (that is, at the apex of the cone formed in the electron energy-momentum relation of quasi-relativistic particles)
    \item recent experiments have measured the signatures of confinement in electrostatically defined graphene quantum dots
\end{itemize}


\noindent \textbf{Introduction}
\\
Friday nights in the laboratory of Andre Geim and Konstantin Novoselov at the University of Manchester were routinely set aside for more playful work. One such speculative idea was to rip layers of carbon from a lump of graphite using sticky tape~\cite{Novoselov2011, Geim2011}. Remarkably, they were able to isolate a single atomic layer of graphite: graphene. Just a few years later in 2010, they were jointly awarded the Nobel Prize in physics for their discovery of (and groundbreaking experiments with) graphene, whose atoms are arranged in a purely two-dimensional honeycomb array resembling chicken wire~\cite{Castro2009, Abergel2010, Peres2010, Sarma2011}.

Of all of the many wonderful properties of the 2-D material graphene, perhaps the most interesting stem from its link to particle physics~\cite{KatsnelsonBook}. Near to the Fermi level the electron bandstructure in graphene is linear,
\begin{equation}
\label{eq0000}
E_\pm = \pm v_{\mathrm F} |\boldsymbol{p}|,
\end{equation}
suggesting its electrons and holes, of momentum $\boldsymbol{p}$, are massless particles which travel with some constant speed $v_{\mathrm F}$ (in some sense, an effective speed-of-light for the material). Furthermore, the Hamiltonian describing these charge carries formally maps onto a massless, two-dimensional Dirac equation such that a treasure trove of quasi-relativistic phenomena arise in a humble pencil trace of graphite~\cite{Shen2012, Thaller2013}. Some examples of such phenomena include Klein tunnelling, atomic collapse, and Lorentz boosts~\cite{Chen2007, Shytov2007, Shytov2009}.

Explicitly, the single electron Hamiltonian $\hat{H}$ describing massless Dirac fermions in two-dimensions $\boldsymbol{r} = (x, y)$ reads 
\begin{equation}
\label{eq1}
\hat{H} = v_{\mathrm F} \boldsymbol \sigma \cdot \boldsymbol{\hat p} + V(\boldsymbol{r}) ,
\end{equation}
where $v_{\mathrm F}$ is the Fermi velocity of the Dirac material, the two-dimensional momentum operators are $\boldsymbol{\hat p} = (\hat p_x, \hat p_y)$, $\boldsymbol \sigma = (\sigma_x, \sigma_y, \sigma_z)$ are Pauli's spin matrices and $V(\boldsymbol{r})$ is a scalar confinement potential. Importantly, the $2 \times 2$ matrix Hamiltonian $\hat{H}$ of Eq.~\eqref{eq1} must act on a two-component spinor wavefunction $\Psi$, where
\begin{equation}
\label{eq2sdsd}
\Psi (\boldsymbol{r}) = 
\left(
 \begin{array}{c}
\psi_A (\boldsymbol{r}) \\
\psi_B (\boldsymbol{r})
 \end{array}
\right),
\end{equation}
which in the case of graphene has arisen from the two interlocking triangular sublattices $A$ and $B$, together creating the overall honeycomb lattice.
\\

\noindent \textbf{Klein tunnelling}
\\
The spinor wavefunction of graphene gives rise to the concept of pseudospin~\cite{Beenakker2008}, which is encapsulated by the relation $\boldsymbol \sigma \cdot \boldsymbol{\hat p}/ |\boldsymbol{p}| = \pm 1$. This neat equations shows that the direction of the charge carrier motion is pinned: for electrons $\boldsymbol \sigma \cdot \boldsymbol{\hat p}/ |\boldsymbol{p}| = +1$ and for holes $\boldsymbol \sigma \cdot \boldsymbol{\hat p}/ |\boldsymbol{p}| = -1$. As such, it is difficult to backscatter massless Dirac particles, which makes them highly desirable for electronic devices requiring significant mobilities.

In nonrelativistic quantum mechanics, a particle impinging upon a one-dimensional potential barrier can tunnel pass through the obstacle despite classical physics suggesting it would be reflected. Nevertheless, the transmission probability decreases exponentially with increasing barrier height. However, in relativistic quantum mechanics an even more counter initiative phenomena can occur. When a relativistic particle meets a large potential barrier it may pass through perfectly, that is with a transmission probability of one~\cite{Allain2011}. This effect is known as Klein tunnelling, and it can be understood as arising due to the conservation of pseudospin~\cite{KatsnelsonKlein2006}. An electron moving to the right becomes a right-moving hole under the high potential barrier and an electron again outside of the barrier, always ensuring $\boldsymbol \sigma \cdot \boldsymbol{\hat p}/ |\boldsymbol{p}| = 1$ throughout the propagation, such that the particle continues its rightwards journey. Clearly, confining massless Dirac particles electrostatically in one dimension presents a nontrivial task due to this tunnelling phenomenon.
\\

\noindent \textbf{Confinement in two-dimensions}
\\
The Hamiltonian $\hat{H}$ of Eq.~\eqref{eq1} acts upon a spinor wavefunction $\Psi$, which has two-components as shown in Eq.~\eqref{eq2sdsd}. When considering radially symmetric problems, such that the scalar potential $V(\boldsymbol{r}) = V(r)$, the wavefunction $\Psi$ in radial coordinates $(r,\theta)$ has the separable form 
\begin{equation}
\label{eq2}
\Psi(r,\theta) = \frac{\mathrm{e}^{ \mathrm{i} m\theta} }{\sqrt{2\pi}} \left(
 \begin{array}{c}
\chi_A(r) \\ \mathrm{i} \mathrm{e}^{ \mathrm{i} \theta}\chi_B(r)
 \end{array}
\right),
\end{equation}
where the two radial functions $\chi_A(r)$ and $\chi_B(r)$ are to be found, and the prefactor $1/\sqrt{2\pi}$ normalizes the angular part of the wavefunction. The integer quantum number $m = 0,\pm1, \pm2, ...$ arises from a consideration of the total angular momentum quantum number $j_z$, where
\begin{equation}
\label{eqsdssfggdd6}
j_z = m +1/2.
\end{equation}
Explicitly, the spinor wavefunction of Eq.~\eqref{eq2} satisfies the eigenvalue equation $\hat{J_z} \Psi = j_z \Psi$, where the angular momentum operator $\hat{J_z}$ reads
\begin{equation}
\label{eqsdssd6}
\hat{J_z} = -\mathrm{i} \partial_{\theta} + \sigma_z/2.
\end{equation}
The wavefunction $\Psi$ given by Eq.~\eqref{eq2} can then be seen to separate the two spatial variables in the Schr\"{o}dinger equation
\begin{equation}
\label{dfdfdf}
\hat{H} \Psi(r,\theta) = E \Psi(r,\theta),
\end{equation}
so that, after using the momentum component relations $\hat p_x = -\mathrm{i} \hbar \partial_x$ and  $\hat p_y = -\mathrm{i} \hbar \partial_y$, as well as the Cartesian to polar coordinate partial derivatives
\begin{equation}
\label{dfddfdfdffdf}
\partial_x = \cos \theta \partial_r - \tfrac{\sin \theta}{r} \partial_\theta, \quad 
\partial_y = \sin \theta \partial_r + \tfrac{\cos \theta}{r} \partial_\theta,
\end{equation}
one arrives at a pair of coupled first-order differential equations for the two radial components $\chi_A(r)$ and $\chi_B(r)$,
\begin{subequations}
\label{eq3}
 \begin{align}
  \left(\partial_r + \tfrac{m+1}{r} \right) \chi_B &= [\varepsilon-U(r)] \chi_A, \label{eq4} \\
  \left(-\partial_r + \tfrac{m}{r}   \right) \chi_A &= [\varepsilon-U(r)] \chi_B. \label{eq5}
 \end{align}
\end{subequations}
Here the scaled eigenvalue $\varepsilon = E/\hbar v_{\mathrm F}$, and the scaled potential function $U(r) = V(r)/\hbar v_{\mathrm F}$, absorb the dependence on the Fermi velocity $v_{\mathrm F}$.

A second-order differential equation for the upper component of the radial wavefunction $\chi_A(r)$ only can be computed by disentangling the system of two equations given by Eq.~\eqref{eq3}, resulting in
\begin{align}
\label{eq6}
\chi_A''& + \left( \tfrac{1}{r} + \tfrac{U'}{\varepsilon-U} \right) \chi_A' \nonumber \\
&+ \left( \left[ \varepsilon-U \right]^2 - \tfrac{m^2}{r^2} - \tfrac{m}{r} \tfrac{U'}{\varepsilon-U} \right) \chi_A = 0.
\end{align}
Here the notation $'$ and $''$ denotes taking one or two derivatives of the function with respect to $r$. A corresponding equation for the lower radial wavefunction component $\chi_B(r)$ may be obtained upon making the replacements $A \to B$ and $m \to -(m+1)$ in Eq.~\eqref{eq6}, as follows from Eq.~\eqref{eq3}.

Importantly, a consideration of any fast-decaying function $U(r)$ allows one to neglect all terms involving both $U$ and $U'$ in Eq.~\eqref{eq6}, such that the equation essentially becomes Bessel's differential equation. The two linearly independent solutions, being Bessel functions of the first and second kinds, are standard scattering solutions and so one is lead to conclude bound states are seemingly not possible, as was pointed out by Tudorovskiy and Chaplik~\cite{Tudorovskiy2007}. However, this argument implicitly assumes the states considered have a nonzero energy, such that $\varepsilon \ne 0$ always in Eq.~\eqref{eq6}. If one allows for zero energy states ($\varepsilon = 0$) then Eq.~\eqref{eq6} no longer maps onto Bessel's differential equation, opening up an opportunity for bound states with electrostatic confining potentials to plausibly emerge~\cite{Bardarson2009}.
\\

\noindent \textbf{Maxwell's fish-eye lens and bound states}
\\
Let us try to construct zero energy ($\varepsilon = 0$) solutions of the massless Dirac equation, associated with the Hamiltonian of Eq.~\eqref{eq1}. Inspired by the Lorentzian function exploited by Maxwell for his fish-eye lens~\cite{MaxwellBook}, we shall choose the smooth confining potential~\cite{Downing2011, Downing2017} 
\begin{equation}
\label{toy1}
V(r) = \frac{-V_0}{1+r^2/d^2},
\end{equation}
where the potential strength is $V_0$, while $d$ measures the spatial extent of the potential. Remarkably, this bell-shaped function allows for an exact solution of Eq.~\eqref{dfdfdf}. 

\begin{figure*}[tb]
 \includegraphics[width=1.0\linewidth]{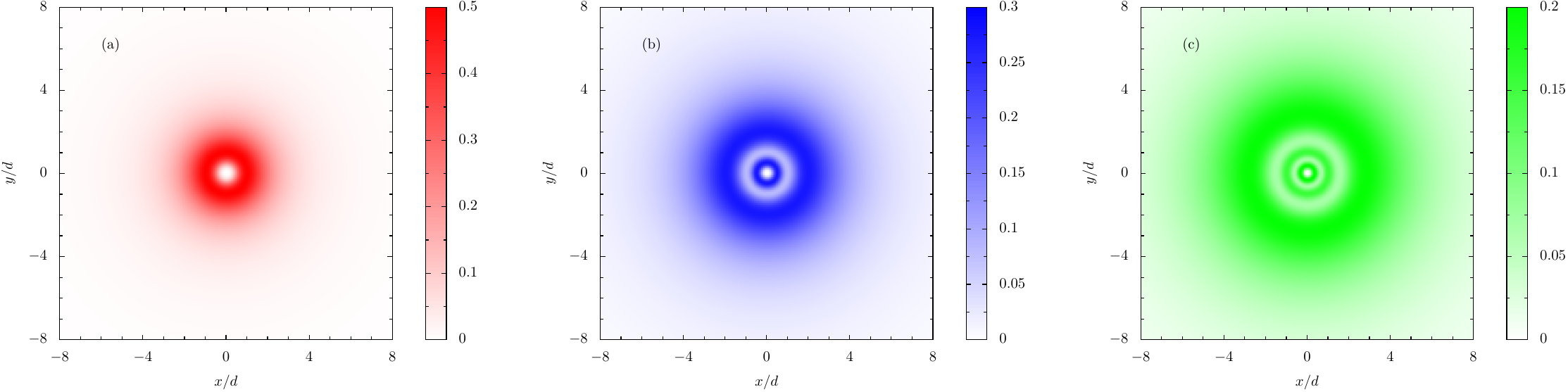}
 \caption{ Probability densities of bound states in a Lorentzian confining potential [cf. Eq.~\eqref{toy6}]. Panels (a, b, c): the quantum number $n = 0, 1, 2$ respectively. In the figure, the quantum number angular momentum $m = 1$.}
 \label{chappy}
\end{figure*}

One may solve the coupled Eq.~\eqref{eq3} by initially finding just the upper radial wavefunction component $\chi_A(r)$. A short-range analysis of Eq.~\eqref{eq6} suggests that at $r \to 0$ the solution should be of the form
\begin{equation}
\label{toysdsdsdsdsd2}
\chi_A \sim r^{|m|},
\end{equation}
where the alternative solution $\chi_A \sim r^{-|m|}$ is disgarded due to its divergence at the origin, as is familiar from non-relativistic central potential problems. A similar long-range analysis as $r \to \infty$ suggests the proper solution should decay like
\begin{equation}
\label{toysdsdsdsdsdsdssd2}
 \chi_A \sim \frac{1}{r^{2 p_m-|m|}}.
\end{equation}
 In the above equation, we have introduced the $m$-dependent parameter $p_m$, defined as
\begin{equation}
\label{toy2}
p_m = \tfrac{1}{2} \left( 1 + |m| + |1+m| \right).
\end{equation}
For later convenience, if we switch to a new variable $\xi$, where
\begin{equation}
\label{toy3sdssd}
\xi = (r/d)^2,
\end{equation}
an appropriate ansatz solution of Eq.~\eqref{eq6}, which accounts for both the short-range and long-range behaviours as encapsulated by Eq.~\eqref{toysdsdsdsdsd2} and Eq.~\eqref{toysdsdsdsdsdsdssd2} respectively, is  
\begin{equation}
\label{toy3}
\chi_A = \frac{C}{d} \frac{\xi^{\tfrac{|m|}{2}}}{(1+\xi)^{p_m}} w(\xi),
\end{equation}
where we have introduced the dimensionless number $C$ as a normalization constant for the radial part of the wavefunction, which ensures that
\begin{equation}
\label{toyfgfgdf3}
 \int_0^\infty \left( |\chi_A|^2 + |\chi_B|^2 \right) r\mathrm{d}r= 1.
\end{equation}
In Eq.~\eqref{toy3}, $w(\xi)$ is an unknown function which has no influence on the small or large asymptotics of the overall upper radial wavefunction component $\chi_A$. Upon substituting Eq.~\eqref{toy3} into Eq.~\eqref{eq6}, we find the following second-order differential equation determining the as yet unspecified function $w(\xi)$ 
\begin{align}
\label{toy4}
 &\xi(1+\xi)^2 w''(\xi) \nonumber \\
 &+ (1+\xi) \left[ 1 + |m| + (2+|m|-2 p_m) \xi \right] w'(\xi) \nonumber \\
 &+ \left[ (\tfrac{V_0 d}{2 \hbar v_{\mathrm F}})^2 - p_m^2 \right] w(\xi) = 0.
\end{align}
The solution of Eq.~\eqref{toy4} can be given in terms of the Gauss hypergeometric function $_2F_1\left( a, b; c; z\right)$, which has the power series definition
\begin{equation}
\label{toy5}
_2F_1\left( a, b; c; z\right) = 1 + \frac{a b}{c} \frac{z}{1!} + \frac{a(a+1) b(b+1)}{c(c+1)} \frac{z^2}{2!} + \ldots
\end{equation}
One finds the explicit form of the solution of Eq.~\eqref{toy4} as
\begin{equation}
\label{toy5}
 w(\xi) =  \,_2F_1\left(  p_m + \tfrac{V_0 d}{2 \hbar v_{\mathrm F}}  , p_m - \tfrac{V_0 d}{2 \hbar v_{\mathrm F}} ;  1 + |m| ; \tfrac{\xi}{1+\xi} \right),
\end{equation}
while the precise structure of $\chi_B$ is now readily obtained from Eq.~\eqref{eq5} since $\chi_A$ has been found.

It is implicit that the infinite power series inside Eq.~\eqref{toy5} must be terminated, so that $\chi_A $ satisfies the required conditions on its limiting behavior, as determined at the outset by Eq.~\eqref{toysdsdsdsdsd2} and Eq.~\eqref{toysdsdsdsdsdsdssd2}. Given this termination, the radial asymptotics of the radial component wavefunctions are
\begin{equation}
\label{toy121}
\lim_{r \to \infty}
\left(
 \begin{array}{c}
\chi_A\\ \chi_B
 \end{array}
\right) \propto \left(
 \begin{array}{c}
1/r^{1+|1+m|} \\ 
1/r^{1+|m|}
 \end{array}
\right), 
\end{equation}
which suggests that the presence of algebraically decaying, but marginally non-square-integrable, extended states with quantum numbers $m=\{0, -1\}$, and otherwise genuine, fully square-integrable bound states with $m \ne \{0, -1\}$. Some typical bound states are plotted in Fig.~\ref{chappy}.

The quantization condition for zero energy ($\epsilon = 0$ or equivalently $E = 0$) bound states arises by assigning the second argument in the hypergeometric function of Eq.~\eqref{toy5} to be a non-positive integer $n$, thus terminating the power series and leading to the bound state conditions
\begin{equation}
\label{toy6}
\frac{V_0 d}{\hbar v_{\mathrm F} } = 2 \left( n + p_m \right), \quad\quad\quad  E = 0,
\end{equation}
where $n = 0, 1, 2 ...$, and the $m$-dependent parameter $p_m$ is defined in Eq.~\eqref{toy2}. All of the modes of Eq.~\eqref{toy6} appear at even integer values of the dimensionless quantity $V_0 d/\hbar v_{\mathrm F}$, which are associated with increasingly large degeneracies. For example, at the bound state condition $V_0 d/\hbar v_{\mathrm F} = 4$ there is a two-fold degeneracy, due to the states $(n, m) = (0, 1)$ and $(0, -2)$, and there is a four-fold degeneracy when $V_0 d/\hbar v_{\mathrm F} = 6$, due to the states $(n, m) = (0, 2)$, $(0, -3)$, $(1, 1)$ and $(1, -2)$.

The features of this fish-eye lens model suggest some general conclusions about bound states of massless Dirac particles may be drawn. Firstly, bound states may only arise at the apex of the energy-momentum cone (the so-called Dirac point) so they are zero-energy bound states ($E = 0$). Secondly, they are only supported at critical values of the dimensionless parameter arising from the potential strength and shape (in this case, $V_0 d/\hbar v_{\mathrm F}$). Thirdly, there is a threshold value at which the first bound state appears (in our example, $V_0 d/\hbar v_{\mathrm F} = 4$). Finally, the angular momentum quantum number $m$ is required to be $m \ge 1$ or $m \le -2$ for the state to be normalizable. States with $m = 0$ and $m = -1$, or equivalently from Eq.~\eqref{eqsdssd6} with total angular momentum number $j_z = |1/2|$, are a marginal kind of extended state.
\\

\noindent \textbf{Experiments}
\\
The quest to observe bound states of massless Dirac particles in graphene and related materials started soon after the isolation of graphene by Geim and Novoselov in Manchester in 2004. Early work used methods like applying high magnetic fields or chemical engineering~\cite{Goerbig2011}, but such techniques are arguably undesirable for future technological applications. Therefore, electrostatic confinement remained a holy grail in the early years of massless Dirac fermion research.

A breakthrough was reported in 2015 by Yue Zhao and co-workers, who were influenced by the so-called whispering gallery modes known from the closed acoustics of cathedrals~\cite{Zhao2015}. The team used a scanning tunneling probe to create a circular p-n junction, and exploited the tunability of the charge carriers of graphene to engineer controllable whispering gallery mode resonators. They observed various degrees of confinement via resonances in the tunneling spectrum, where particularly strong confinement was found by carefully tuning the cavity radius using gate potentials. Therefore, this pioneering work provided a clear perspective for the creation of zero-energy bound states electrostatically.

Soon afterwards, Juwon Lee and colleagues observed bound states in graphene quantum dots in a landmark paper from 2016~\cite{Lee2016}. They fabricated circular quantum dots by careful manipulations of defect charges in the insulator substrate sitting below the monolayer of graphene. The team was able to map spatially the electronic structure inside and outside the quantum dot, providing insight into the quantum interference patterns associated with longer living states.

In the same year, Christopher Gutierrez and co-workers similarly used substrate engineering to create a circular graphene quantum dot~\cite{Gutierrez2016}. They revealed long-lived states locally pinned within a larger monolayer sheet of graphene. In particular, they observed essentially bound states, at particular energies and of a certain angular momentum, which were achieved by carefully tuning the geometry of the barrier.

Taken together, these three aforementioned experimental works suggest that the creation and mastery of electrostatic confinement with massless Dirac particles, with a view towards future exploitation in nanoscopic devices, remains an important area of condensed matter physics research.
\\

\noindent \textbf{Summary and future directions}
\\
We have discussed how Dirac materials, such as the Nobel Prize winning wonder material graphene, allow one to explore some aspects of quantum electrodynamics in condensed matter physics thanks to their shared reliance on various forms of Dirac equation. One such example is Klein tunnelling, or the perfect tunnelling of massless particles normally incident on high potential barriers. This effect suggests that the creation of electrostatic quantum dots in Dirac materials is a highly nontrivial task.

We have considered theoretically a possible route to create trapping of massless particles, by considering zero energy states in electrostatic confining potentials. Via a beautiful, exactly-solvable model, inspired by Maxwell's fish-eye lens, we have described some key properties of such novel bound states arising at the Dirac point.

Finally, we have briefly reviewed some exciting experimental work on graphene quantum dots. As it stands, the complete mastery of the confinement of massless particles, a key blockage preventing the development of ultrafast, ultracompact electronics based on Dirac materials, remains fertile ground for modern condensed matter physics studies.
\\

\noindent \textbf{Acknowledgments}
\\
CAD is supported by the Royal Society via a University Research Fellowship (URF\slash R1\slash 201158). MEP is supported by the EU H2020-MSCA RISE projects TERASSE (Project No. 823878) and DiSeTCom (Project No.823728).
\\




\begin{thebibliography}{100}




\bibitem[Abergel~et~al.,~2010]
{Abergel2010}
Abergel, D. S. L. et al., 2010.
Properties of Graphene: A Theoretical Perspective,
\href{https://doi.org/10.1080/00018732.2010.487978}
{Adv. Phys. \textbf{59}, 261}.


\bibitem[Allain~and~Fuchs,~2011]
{Allain2011}
Allain, P. E., Fuchs, J. N., 2011.
Klein tunneling in graphene: optics with massless electrons,
\href{https://doi.org/10.1140/epjb/e2011-20351-3}
{Eur. Phys. J. B \textbf{83}, 301}.

\bibitem[Bardarson~et~al.,~2009]
{Bardarson2009}
Bardarson, J. H. et al., 2009.
Electrostatic confinement of electrons in an integrable graphene quantum dot,
\href{https://doi.org/10.1103/PhysRevLett.102.226803}
{Phys. Rev. Lett. \textbf{102}, 226803}.

\bibitem[Beenakker,~2008]
{Beenakker2008}
Beenakker, C. W. J., 2008.
Colloquium: Andreev reflection and Klein tunneling in graphene,
\href{https://doi.org/10.1103/RevModPhys.80.1337}
{Rev. Mod. Phys. \textbf{80}, 1337}.

\bibitem[Castro~Neto~et~al.,~2009]
{Castro2009}
Castro Neto, A. H. et al., 2009.
The electronic properties of graphene,
\href{https://doi.org/10.1103/RevModPhys.83.851}
{Rev. Mod. Phys. \textbf{81}, 109}.


\bibitem[Chen~et~al.,~2007]
{Chen2007}
Chen, H.-Y. et al., 2007.
Fock-Darwin states of Dirac electrons in graphene-based artificial atoms,
\href{https://doi.org/10.1103/PhysRevLett.98.186803}
{Phys. Rev. Lett. \textbf{98}, 186803}.

\bibitem[Das~Sarma~et~al.,~2011]
{Sarma2011}
Das Sarma, S. et al., 2011.
Electronic transport in two-dimensional graphene,
\href{https://doi.org/10.1103/RevModPhys.83.407}
{Rev. Mod. Phys. \textbf{83}, 407}.

\bibitem[Downing~et~al.,~2011]
{Downing2011}
Downing, C. A. et al., 2011.
Zero-energy states in graphene quantum dots and rings,
\href{https://doi.org/10.1103/PhysRevB.84.155437}
{Phys. Rev. B \textbf{84}, 155437}.

\bibitem[Downing~et~al.,~2017]
{Downing2017}
Downing, C. A., Portnoi, M. E., 2017.
Bielectron vortices in two-dimensional Dirac semimetals,
\href{https://doi.org/10.1038/s41467-017-00949-y}
{Nat. Commun. \textbf{8}, 897}.

\bibitem[Geim,~2011]
{Geim2011}
Geim, A. K., 2011.
Nobel lecture: Random walk to graphene,
\href{https://doi.org/10.1103/RevModPhys.83.851}
{Rev. Mod. Phys. \textbf{83}, 851}.

\bibitem[Goerbig~et~al.,~2011]
{Goerbig2011}
Goerbig, M. O., 2011.
Electronic properties of graphene in a strong magnetic field,
\href{https://doi.org/10.1103/RevModPhys.83.1193}
{Rev. Mod. Phys. \textbf{83}, 1193}.

\bibitem[Gutierrez~et~al.,~2016]
{Gutierrez2016}
Gutierrez, C. et al., 2016.
Klein tunnelling and electron trapping in nanometre-scale graphene quantum dots,
\href{https://doi.org/10.1038/nphys3806}
{Nat. Phys. \textbf{12}, 1069}.

\bibitem[Katsnelson~et~al.,~2006]
{KatsnelsonKlein2006}
Katsnelson, M., Novoselov, K., Geim, A., 2006.
Chiral tunnelling and the Klein paradox in graphene,
\href{https://doi.org/10.1038/nphys384}
{Nature Phys. \textbf{2}, 620}.

\bibitem[Katsnelson,~2012]
{KatsnelsonBook}
Katsnelson, M. I., 2012.
\textit{Graphene: Carbon in Two Dimensions}
(Cambridge University Press).


\bibitem[Lee~et~al.,~2016]
{Lee2016}
Lee, J. et al., 2016.
Imaging electrostatically confined Dirac fermions in graphene quantum dots,
\href{https://doi.org/10.1038/nphys3805}
{Nat. Phys. \textbf{12}, 1032}.


\bibitem[Niven,~1890]
{MaxwellBook}
Niven, W. D., 1890.
\textit{The Scientific Papers of James Clerk Maxwell: Edited by WD Niven}
(Dover Publications).

\bibitem[Novoselov,~2011]
{Novoselov2011}
Novoselov, K. S., 2011.
Nobel lecture: Graphene: materials in the flatland,
\href{https://doi.org/10.1103/RevModPhys.83.837}
{Rev. Mod. Phys. \textbf{83}, 837}.

\bibitem[Peres~et~al.,~2010]
{Peres2010}
Peres, N. M. R. et al., 2010.
Colloquium: the transport properties of graphene: an introduction,
\href{https://doi.org/10.1103/RevModPhys.82.2673}
{Rev. Mod. Phys. \textbf{82}, 2673}.




\bibitem[Shen,~2012]
{Shen2012}
Shen, S.-Q., 2012.
\textit{Topological Insulators: Dirac equation in Condensed Matters}
(Springer Berlin Heidelberg).



\bibitem[Shytov~et~al.,~2007]
{Shytov2007}
Shytov, A. V. et al., 2007.
Atomic collapse and quasi–Rydberg states in graphene,
\href{https://doi.org/10.1103/PhysRevLett.99.246802}
{Phys. Rev. Lett. \textbf{99}, 246802}.

\bibitem[Shytov~et~al.,~2009]
{Shytov2009}
Shytov, A. V. et al., 2009.
Atomic collapse, Lorentz boosts, Klein scattering, and other quantum-relativistic phenomena in graphene,
\href{https://doi.org/10.1016/j.ssc.2009.02.043}
{Solid State Commun. \textbf{149}, 1087}.

\bibitem[Thaller,~2013]
{Thaller2013}
Thaller, B., 2013.
\textit{The Dirac equation}
(Springer Berlin Heidelberg).





\bibitem[Tudorovskiy~and~Chaplik~2007]
{Tudorovskiy2007}
Tudorovskiy, T. Ya., Chaplik, A. V., 2007.
Spatially inhomogeneous states of charge carriers in graphene,
\href{https://doi.org/10.1134/S002136400623010X}
{JETP Lett. \textbf{84}, 619}.



\bibitem[Zhao~et~al.,~2015]
{Zhao2015}
Zhao, Y. et al., 2015.
Creating and probing electron whispering-gallery modes in graphene,
\href{https://doi.org/10.1126/science.aaa7469}
{Science \textbf{348}, 672}.



\end{thebibliography}
\end{document}